%% file: main.tex
\documentclass[lettersize,journal]{IEEEtran}
\usepackage{amsmath,amsfonts}
\usepackage{algorithmic}
\usepackage{array}
\usepackage[caption=false,font=normalsize,labelfont=sf,textfont=sf]{subfig}
\usepackage{textcomp}
\usepackage{stfloats}
\usepackage{url}
\usepackage{verbatim}
\usepackage{graphicx}
\usepackage{adjustbox}
\usepackage{pifont}
\usepackage{multirow}
\def\BibTeX{{\rm B\kern-.05em{\sc i\kern-.025em b}\kern-.08em
    T\kern-.1667em\lower.7ex\hbox{E}\kern-.125emX}}
\usepackage{balance}
\usepackage{xspace}
\usepackage{xcolor}

\newcommand{\cmark}{\ding{51}}
\newcommand{\xmark}{\ding{55}}
\usepackage{booktabs}
\usepackage{graphicx}
\usepackage{longtable}
\usepackage{makecell}
\usepackage{supertabular}
\usepackage{wrapfig}
\usepackage{amsmath}
\newcommand{\model}{{AnyECG}\xspace}
\newcommand{\tokenizer}{{ECG Tokenizer}\xspace}

\begin{document}
% \title{AnyECG: Foundational Models for Electrocardiogram Analysis}
\title{AnyECG: Foundational Models for Multitask Cardiac Analysis in Real-World Settings}
\author{Yue Wang$^{1,*}$, Xu Cao$^{2,*}$, Yaojun Hu$^{1}$, Haochao Ying$^{7}$, Hongxia Xu$^{1}$, Ruijia Wu$^{3}$ \\ James Matthew Rehg$^{4}$, 
Jimeng Sun$^{6}$, Jian Wu$^{1,7}$, and Jintai Chen$^{5,\dagger}$%
\thanks{*Equal contribution.}%

% \thanks{
% This research was supported by the National Natural Science Foundation of China under Grant No.~62476246, the Zhejiang Provincial Natural Science Foundation of China under Grant No.~LY23F020019, and the Opening Foundation of the State Key Laboratory of Transvascular Implantation Devices under Grant No.~SKLTID2024003.}
% \thanks{
% This research was supported by the National Natural Science Foundation
% of China under Grant No. 82202984,  No. 62476246, the Zhejiang Provincial Natural Science
% Foundation of China under Grant No. LY23F020019, and the Opening
% Foundation of the State Key Laboratory of Transvascular Implantation Devices
% under Grant No. SKLTID2024003.
% Jintai Chen is partially supported by the internal fund from the Hong
% Kong University of Science and Technology (Guangzhou).}
\thanks{$^{1}$ The College of Computer Science and Technology, Zhejiang University, Hangzhou 310012, China, is also associated with the State Key Laboratory of Transvascular Implantation Devices of the Second Affiliated Hospital, Zhejiang University School of Medicine, Hangzhou 310009, China, and the Zhejiang Key Laboratory of Medical Imaging Artificial Intelligence, Hangzhou 310058, China. Email: \texttt{ywang2022@zju.edu.cn}}%
% \thanks{$^{2}$College of Computer Science and Technology, Zhejiang University, Hangzhou, 310012, China.}%
\thanks{$^{2}$Department of Computer Science, University of Illinois at Urbana-Champaign, Urbana, Illinois, USA. Email: \texttt{xucao2@illinois.edu}}%
\thanks{$^{3}$Antai College of Economics and Management, Shanghai Jiao Tong University, Shanghai, China}%
\thanks{$^{4}$School of Computer Science, University of Illinois at Urbana-Champaign, Urbana, Illinois, USA.}%
\thanks{$^{5}$Information Hub, Hong Kong University of Science and Technology (Guangzhou), Guangzhou, 511458, China.  }%
\thanks{$^{6}$University of Illinois at Urbana-Champaign, Urbana, Illinois, USA.}%
\thanks{$^{7}$ the School of Public Health and Second Affiliated Hospital, Zhejiang University School of Medicine, Hangzhou 310058, China. }
\thanks{$^{\dagger}$Corresponding author. E-mail:\texttt{jtchen721@gmail.com}}%
}

% \markboth{Journal of \LaTeX\ Class Files,~Vol.~18, No.~9, September~2020}%
% {How to Use the IEEEtran \LaTeX \ Templates}

\maketitle

\begin{abstract}
Electrocardiogram (ECG), a non-invasive and affordable tool for cardiac monitoring, is highly sensitive in detecting acute heart attacks. However, due to the lengthy nature of ECG recordings, numerous machine learning methods have been developed for automated heart disease detection to reduce human workload. Despite these efforts, performance remains suboptimal. A key obstacle is the inherent complexity of ECG data, which includes heterogeneity (\textit{e.g.}, varying sampling rates), high levels of noise, demographic-related pattern shifts, and intricate rhythm-event associations.
To overcome these challenges, this paper introduces {\model}, a foundational model designed to extract robust representations from any real-world ECG data. Specifically, a tailored {\tokenizer} encodes each fixed-duration ECG fragment into a token and, guided by proxy tasks, converts noisy, continuous ECG features into discrete, compact, and clinically meaningful local \texttt{rhythm codes}. These codes encapsulate basic morphological, frequency, and demographic information (\textit{e.g.}, sex), effectively mitigating signal noise. We further pre-train the {\model} to learn rhythmic pattern associations across ECG tokens, enabling the capture of cardiac event semantics.
By being jointly pre-trained on diverse ECG data sources, {\model} is capable of generalizing across a wide range of downstream tasks where ECG signals are recorded from various devices and scenarios. The experimental results show that AnyECG achieves an average performance improvement of 6\% across four critical tasks—anomaly detection, arrhythmia classification, corrupted lead generation, and ultra-long ECG recognition. {\model} learns common ECG rhythm from data and significantly outperforms state-of-the-art methods in each of these tasks.

% Experimental results demonstrate significant improvements over state-of-the-art methods, with AnyECG achieving an average performance gain of 6\% across four critical tasks—anomaly detection, arrhythmia classification, corrupted lead generation, and ultra-long ECG analysis.  Experimental results in anomaly detection, arrhythmia detection, corrupted lead generation, and ultra-long ECG signal analysis demonstrate that {\model} learns common ECG knowledge from data and significantly outperforms cutting-edge methods in each respective task.
\end{abstract}


% \begin{document}
% \input{sec/1.abstract}
\input{sec/2.introduction}
\input{sec/3.related_work_new}

\input{sec/4.methodology_new}
\input{sec/5.experiment}

\input{sec/7.discussion}
\input{sec/6.conclusion}

\bibliographystyle{IEEEtran}
\bibliography{paper}
% \bibliographystyle{paper}
% \newpage
% \clearpage
% \appendices

\section{Appendix}
% \section{Appendix}
% \input{sec/content}
% \input{sec/3.related_work}

\input{sec/appendix}

\end{document}

%% file: sec/2.introduction.tex
\section{Introduction}

The electrocardiogram (ECG) is a widely used test that records the heart's electrical activity, facilitating the monitoring and diagnosis of various cardiac conditions. Due to variations in ECG devices, recording conditions, patient characteristics, the length of recorded ECG signals, the number of leads, the sampling rates, as well as the signal-to-noise ratio (SNR), can vary significantly. For example, in non-clinical settings, wearable devices typically collect long-term single-lead or dual-lead ECG signals at lower sampling rates, covering a variety of human activity scenarios, which often results in higher noise levels~\cite{abbaspourazad2023large,ansari2023deep}. In contrast, standard devices used in hospital outpatient clinics capture high-resolution eight-, twelve-, or eighteen-lead ECG signals in a resting state for diagnostic purposes~\cite{herman2024validation}. Additionally, the noise in ECG data can originate from device artifacts, baseline wander, muscle noise, as well as external interference~\cite{singh2022attention}.
% , which not only complicates data analysis but also increases the risk of spurious correlations affecting the identification of disease manifestations. 
These heterogeneity and complexity present major challenges in developing a unified model that can effectively handle ECG signals recorded across various devices, scenarios, and clinical purposes.

Sequence models, such as large language models, developed using large-scale data in the wild have shown significant advantages in learning robust representations and demonstrated robustness in downstream tasks. However, adapting sequence model pre-training approaches to ECG data in real-world settings poses unique challenges: 
\textbf{(1) Heterogeneity:} Real-world ECG signals exhibit significant variations in length (seconds to hours), sampling rates (100--1000~Hz), and channel numbers (1--12 leads) across acquisition devices. Conventional approaches struggle to reconcile these differences: sampling rate discrepancies fundamentally distort temporal patterns (\textit{e.g.}, a heartbeat spans 4$\times$ fewer samples at 125~Hz vs.\ 500~Hz), while unifying high-resolution long-term signals (\textit{e.g.}, \textgreater86M samples for 24-hour 1000~Hz data) exceeds computational limits of standard architectures. Existing methods either ignore these critical variations or impose aggressive downsampling that degrades diagnostically vital features.
% \textbf{(1) Heterogeneity:} real-world ECG signals vary in length, sampling rate, and the number of channels due to differences in devices and scenarios. A unified model (\textit{e.g.}, with fixed tokenizer settings) is needed to effectively manage this diversity while maintaining temporal resolution and avoiding the introduction of artifacts.
% [tech difficulties]
% making it difficult to align and integrate data from multiple sources.
% \textbf{(2) Low SNR:} ECG signals inherently have a low SNR, and pathological waveforms are often subtle, making it easy for noise to interfere with the understanding of critical features.
\textbf{(2) Low SNR:}  ECG signals inherently have a low SNR, and pathological waveforms are often subtle, making it easy for noise to interfere with the understanding of critical features. Meanwhile, during signal extraction, deep learning models tend to prioritize the extraction of high-frequency signals (such as the QRS complex) while neglecting low-frequency features (such as the ST-T segment), leading to an imbalance in the model's spectral sensitivity.

% This noise can obscure important features in the ECG, making it difficult for models to learn accurate representations of the underlying cardiac activity. Models may learn to recognize noise patterns specific to certain devices or settings rather than the true physiological signals, leading to poor generalization.  
\textbf{(3) Demographic shift:} ECG waveforms can vary due to patient demographics (\textit{e.g.}, age, sex, ethnicity). For instance, pediatric ECGs differ from adults in disease presentation and heart rate~\cite{chen2024congenital}, and distinct ethnic groups may exhibit unique ECG characteristics~\cite{jain2010diagnostic}, which hinder models from generalizing across diverse populations. 

To overcome these challenges, we introduce {\model}s, a family of ECG foundational models designed for robust representation learning on ECG signals in diverse, real-world settings. The development of \model involves two main pre-training phases: the Rhythm Quantizer pre-training and the entire {\model} foundation model pre-training. The first phase captures key local rhythmic patterns from noisy ECG signals, while the second learns associations across the rhythmic patterns that implies cardiac events. 
Specifically, we first pad any ECG signals to unified length, segment a signal into a collection of fixed-duration fragments, project each fragment into a token orderly, and pad missing channels, standardizing the diversity in sampling rates, lengths, and channel numbers. We further design a new hierarchical modeling approach to tackle ultra-long ECG signals. In the Rhythm Quantizer training phase, a \texttt{Rhythm Codebook} is established to capture the key local morphological and frequency features inherent in ECG signals.
The {\tokenizer} extracts ECG features that are closely aligned with these \texttt{Rhythm Codes}, effectively reducing noise by matching the input patches to those representative codes. Additionally, the extracted ECG features are also required to recover demographic information about the patient. Then, we apply `masked modeling' approach in the \model pre-training phase, where the model predicts \texttt{Rhythm Code} indices to fill in masked patches. This approach encourages the recovery of masked ECG patches based on their relationship with unmasked ECG patches, enabling the model to learn cardiac event semantics that are essential for downstream tasks.
With these designs, \model can facilitate knowledge transfer across various ECG sources in the wild, enabling to learn shared ECG and cardiac event knowledge that is applicable to downstream tasks. Our contributions are listed below.

\begin{itemize}
    % \item \textbf{Harmonization of Heterogeneous ECG Data:} We present a foundational ECG model that effectively addresses data heterogeneity arising from variations in machine parameters, sampling frequencies, patient characteristics, environmental noise, lead numbers, and lead placement positions, offering broad applicability across ECG analysis tasks.

    \item \textbf{Cardio-Sparse Attention:}Inspired by physicians' prioritization of key cardiac events, we design Cardio-Sparse Attention (CSA), a mechanism that focuses on critical rhythm phases while masking non-informative segments. By mimicking clinicians' diagnostic focus on actionable patterns and ignoring redundant noise, this design reduces computational costs, enabling efficient analysis of ultra-long ECG recordings.

    \item \textbf{Rhythm Quantizer:} We propose a Rhythm Quantizer framework that transforms raw ECG signals into discrete, noise-resilient \texttt{Rhythm Codes} through a multi-view decoding process. By jointly reconstructing morphological waveforms (e.g., QRS complexes), time-frequency patterns (via wavelet decomposition), and demographic attributes (e.g., age-related T-wave variations), this framework explicitly separates pathological features from artifacts. The generated codes provide a unified representation for heterogeneous ECG data, enabling robust analysis in low-SNR scenarios.
    
    % \item \textbf{ECG Foundational Model:} We introduce \model, a foundational ECG model that unifies representation learning by capturing important local rhythm patterns in ECG signals and their semantic relationships, providing a flexible framework adaptable to any ECG signals for various downstream applications.

    % We introduce a new tokenizer that guides the encoding process to focus on ECG morphological features, signal frequency characteristics, and patient attributes, converting ECG signals into compact, noise-resilient representations.

    % \item \textbf{Various Downstream Tasks Adaptability:} By pretrained to learn the associations among ECG tokens, our \model is sensitive to potential cardiac events, demonstrating strong generalization capabilities across various downstream tasks, including anomaly detection, arrhythmia detection, corrupted lead generation, and ultra-long ECG signal analysis.
    \item \textbf{
Foundation Model for Unified Cardiac Diagnostics:} By pretrained to learn the associations among ECG tokens, our \model is sensitive to potential cardiac events and achieves state-of-the-art performance including anomaly detection, arrhythmia detection, corrupted lead generation, and ultra-long ECG signal analysis.  The two-stage pre-training paradigm ensures adaptability to fragmented clinical workflows, providing a unified solution for real-world ECG diagnostics.
    
    % model consistently outperforms state-of-the-art methods across various downstream tasks, confirming its strong generalization capabilities and robustness in managing variability, noise, and diversity in real-world ECG data.
    
    % Extensive experiments show that our model consistently outperforms state-of-the-art methods across various downstream tasks, confirming its strong generalization capabilities and robustness in ECG analysis.
\end{itemize}

%% file: sec/3.related_work_new.tex
\section{Related Works}

\paragraph{Heterogeneous ECG Signal Analysis and Classification}
The application of deep learning techniques has significantly advanced the analysis and classification of ECG signals. However, the heterogeneity of ECG data poses a major challenge for model generalization; models trained on one dataset often do not perform well on others. Consequently, researchers have focused on designing specialized models tailored to specific datasets, employing architectures such as convolutional neural networks (CNNs)~\cite{prathipati2023combining, kucukseymen2022left}, recurrent neural networks (RNNs)~\cite{kumar2023investigation, din2024ecg}, and transformer-based models~\cite{shah2024ecg, ji2024msgformer}. While these efforts have led to incremental performance improvements \cite{srivastava2023apneanet, jasvitha20241d, ribeiro2020automatic, gao2021end}, the gains are often not statistically significant due to the limited size and scope of the datasets used. The absence of a unified model capable of handling the diverse nature of ECG data underscores the need for new approaches that can provide more substantial and broadly applicable performance improvements.

\paragraph{Self-supervised ECG Representation Learning}

Self-supervised learning has emerged as a promising approach for extracting representations from unlabeled ECG signals, enabling the use of large amounts of raw data without manual annotations. Methods such as signal reconstruction, contrastive learning, and masked signal modeling have been explored \cite{yun2024automatic, wu2024improving, li2024learning}. However, existing self-supervised learning methods often struggle to generalize across heterogeneous ECG datasets, especially when faced with varying lead configurations and noise levels. For example, contrastive methods \cite{kiyasseh2021clocs, wang2023adversarial} encourage similar representations for compatible signal segments but do not adequately account for variability introduced by different lead setups. Moreover, the low SNR inherent in ECG data can cause models to focus on reconstructing noisy or redundant signal components due to high correlations among leads, rather than capturing critical physiological information. Models like contrastive predictive coding (CPC)~\cite{mehari2022self} and masked autoencoders~\cite{zhang2022maefe, na2024guiding} often inadvertently emphasize less relevant features, diminishing their effectiveness in capturing essential signal characteristics. This focus on less informative aspects can limit the models' ability to extract meaningful representations that transfer effectively to unseen data or datasets with different characteristics.

%% file: sec/4.methodology_new.tex
\section{Methodology}

\begin{figure*}[t]
    \centering
    \includegraphics[width=0.99\textwidth]{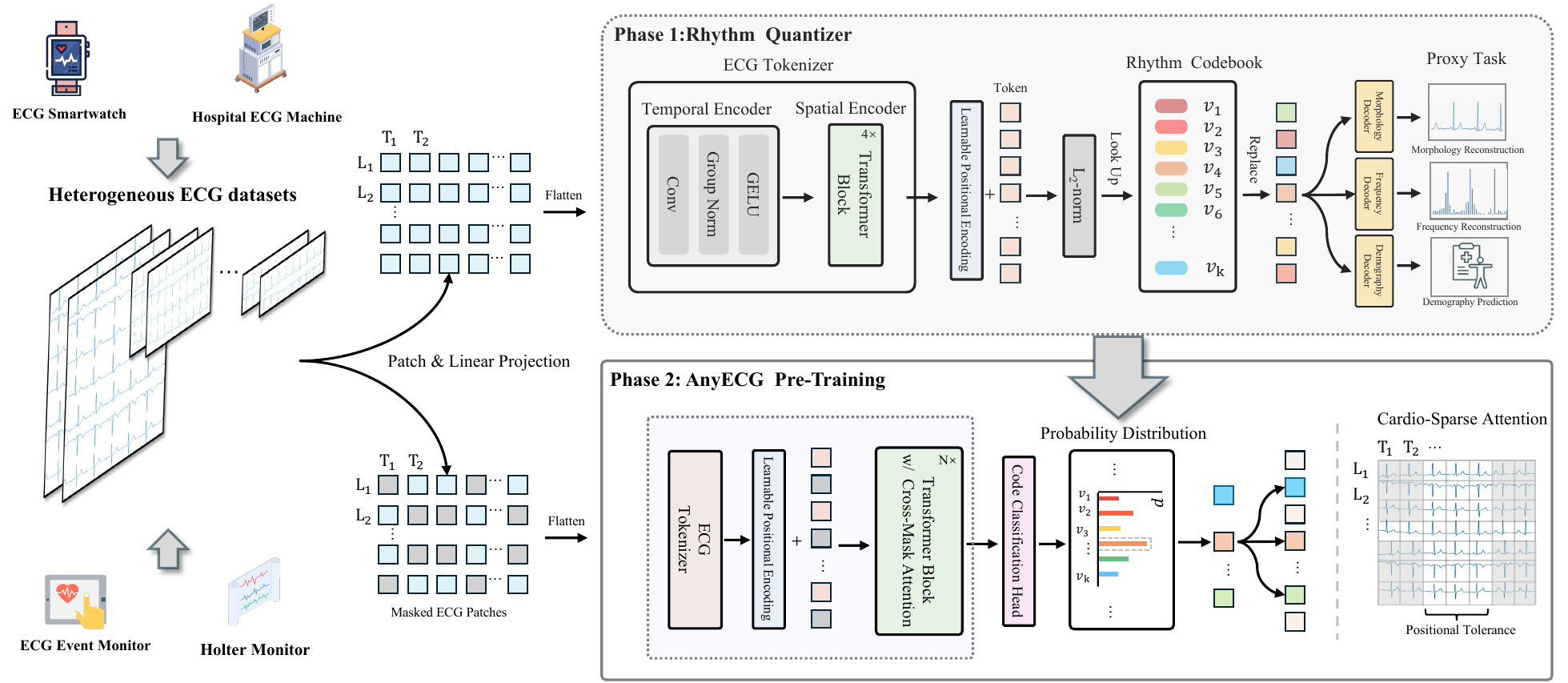}
    \caption{\textbf{Overall architecture and pre-training pipeline of \model.} 
    \model is pre-trained in two steps. The Rhythm Quantizer is pre-trained through proxy tasks to embed morphology, frequency, and demography into tokens \textbf{(up)}. Then, the entire \model, along with the ECG Tokenizer, is further pre-trained by predicting the code indices of the masked tokens to learn the semantic relationships between tokens \textbf{(bottom-left)}. The Cardio-Sparse approach restricts interactions of patches from the same lead or from the same position across different leads \textbf{(bottom-right)}. LN: LayerNorm, Conv: 1D convolution with kernel size of 15.
    % Additionally, a complementary masking approach was proposed to provide more masking perspectives focusing on both heavily masked and minimally masked regions of the ECG signals.
    }
    \label{fig:main}
\end{figure*}
% Our proposed \model adopts the Transformer architecture, incorporating a novel attention module and a special tokenizer that can be adapted to both the self-supervised learning pretraining pipeline and various downstream tasks for any ECG signals, as shown in Figure \ref{fig:main}. The self-supervised learning process of AnyECG is divided into two phases: pretraining rhythm quantizer and pretraining the AnyECG backbone. In Section III-A, we introduce the architecture of AnyECG, detailing its components and design principles. In Sections III-B and III-C, we describe the Rhythm Quantizer and the pre-training process of AnyECG. Finally, in Section III-D, we summarize the key takeaways.

This section presents our proposed \model, which is designed to be adaptable to both the self-supervised learning pre-training pipeline and various downstream tasks for any ECG signals, as illustrated in Figure \ref{fig:main}. 
% The self-supervised learning process of AnyECG is divided into two phases: pre-training rhythm quantizer and pretraining the AnyECG backbone.
In Section III-A, we introduce the architecture of AnyECG, detailing its components and design principles. In Sections III-B and III-C, we describe the Rhythm Quantizer and the pre-training process of AnyECG. Finally, in Section III-D, we summarize the key takeaways.

\subsection{\model Architecture}
\paragraph{ECG Signal Pre-Processing}
To preserve the natural characteristics of the ECG signals, we applied minimal pre-processing steps aimed at reducing noise while maintaining signal integrity. First, we used a bandpass filter between 0.1 Hz and 75 Hz to remove low-frequency noise, followed by a notch filter at 50 Hz to eliminate power-line interference. The ECG signals were then resampled to 300 Hz to standardize the sampling rate across all data sources, as 300 Hz is considered sufficient for diagnosing most cardiac conditions based on the Nyquist-Shannon sampling theorem. Finally, wavelet-based denoising was performed using the `db6' wavelet, following~\cite{ma2022effective}. 
To align with the Transformer input format, a multi-channel ECG signal \( X \in \mathbb{R}^{L \times T} \) (where \( L \) represents the number of ECG leads (channels), and \( T \) denotes the total number of recorded temporal points) was segmented along the time axis into fixed-duration patches of size \( s \). This divides each lead into \( P \) patches, where \( P \) is the minimal positive integer satisfying \( P \times s \geq T \). 
If \( T \) is not divisible by  \( s \), we pad the signal with zeros at the end to reach a length of \( P s \). 
Each patch \( x_{j,k} \in \mathbb{R}^{s} \) is defined as \( x_{j,k} = {X}_{j, \,(k-1)s + 1 : k s} \), where \( j = 1, 2, \dotsc, L \) denotes the lead index and \( k = 1, 2, \dotsc, P \) denotes the patch index along the time axis. 
The total number of patches is \( N = L \times P \), and we flatten them into a patch sequence \( X' \in \mathbb{R}^{N\times s} \) of length \( N \) before feeding it into the model.

\paragraph{Rhythm Quantizer}
The objective of Rhythm Quantizer is to effectively capture both the temporal and spatial features of ECG signals and generate an embedding $H \in \mathbb{R}^{N \times d}$ from \( X'\). We firstly use a temporal encoder to learn local temporal patterns by independently processing each ECG patch $x_{j,k}$. The temporal encoder consists of one 1D convolution, a group normalization, and a GELU activation function. Then, a spatial encoder with 4 Transformer blocks is used. Each patch \( x_{j,k} \) is processed by the temporal encoder and spatial encoder sequentially to obtain an embedding \( h'_{j,k} \in \mathbb{R}^d \). To enhance the model's understanding of the temporal sequence and leads relationships, learnable temporal position encoding \( \tau_k \in \mathbb{R}^d \) and lead position encoding \( \sigma_j \in \mathbb{R}^d \) are along the temporal dimension \( k \) and channel dimension \( j \), separately. The output of the \tokenizer is the temporal-spatial encoding ${h}_{j,k} = h'_{j,k} + \tau_k + \sigma_j$.

\paragraph{Cardio-Sparse Attention (CSA)} Unlike other sequential data like text, ECG signals typically include multiple leads, with the signals at the same positions across leads providing complementary information~\cite{chenelectrocardio}. Therefore, in contrast to conventional multi-head self-attention, we introduce CSA, which differentiates the structure of our \model. CSA allows each patch to interact only with patches within relevant channels (\textit{i.e.} leads) and temporal contexts. We apply CSA as the attention module within the Transformer blocks of both the \tokenizer and \model backbone. Given input $H$, $Q = \text{LayerNorm}({H}) W_Q$, $ K = \text{LayerNorm}({H}) W_K$, $V = {H} W_V$, where \( W_Q, W_K, W_V \in \mathbb{R}^{d \times d_{\text{model}}} \) are learnable projection matrices, \( \text{LayerNorm}(\cdot) \) denotes layer normalization, and \( d_{\text{model}} \) is the model dimension. The CSA is computed as:

% \begin{equation}
% \begin{aligned}
% \text{CSA}(Q, K, V) = \text{softmax}\left( \frac{Q K^\top + M}{\sqrt{d_{\text{head}}}} \right) V
% \end{aligned}
% ,
% \begin{aligned}
% M_{i,j} = \begin{cases}
% 0, & \text{if } j \in \mathcal{A}(i) \\
% -\infty, & \text{otherwise}
% \end{cases}
% \end{aligned}
% \end{equation}

\begin{equation}
\begin{aligned}
\text{CSA}(Q, K, V) &= \text{softmax}\left( \frac{Q K^\top + M}{\sqrt{d_{\text{head}}}} \right) V, \\
M_{i,j} &=
\begin{cases}
0, & \text{if } j \in \mathcal{A}(i), \\
-\infty, & \text{otherwise}.
\end{cases}
\end{aligned}
\end{equation}

where, $d_{head}$ is the number of attention head; \( M \in \mathbb{R}^{N \times N} \) is the attention mask matrix. \( \mathcal{A}(i) \) includes patches from the same lead \( j \) or the same position, as illustrated in Figure~\ref{fig:main} bottom right. Notably, a positional tolerance (mask width) is used to improve the model's robustness, accounting for slight delays in certain leads caused by variations in cardiac signal conduction, which is particularly significant for some diseases. In \model, we adopt the multi-head attention version.

\subsection{Rhythm Quantizer Pre-training}

\subsubsection{\tokenizer with Rhythm Codebook}

\textbf{Motivation.} ECG signals are inherently high-dimensional time-series data, often characterized by a low SNR due to sparse key information and contamination from various types of noise. To address these issues, we propose a vector-quantized rhythm codebook that transforms raw ECG signals into compact, discrete tokens, enabling robust and noise-resistant representation learning. The transformation of  rhythm codebook enhances low-SNR signals into a high-SNR representation, accurately capturing true cardiac activity while minimizing the effects of noise.

Initially, each patch \( x_{j,k} \in \mathbb{R}^s \) represents a portion of the signal over \( w \) time steps in lead \( j \). The tokenizer processes these patches into feature representations, yielding embeddings \( {h}_{j,k}  \in \mathbb{R}^d \), where \( d \) is the dimension of each embedding.
To discretize these continuous embeddings into tokens suitable for subsequent processing, we employ a quantizer that maps each embedding \( h_{j,k} \) to the nearest codeword in a predefined codebook \( V \). The codebook \( V \in \mathbb{R}^{K \times d} \) consists of \( K \) codes \( v_1, v_2, \dotsc, v_K \). The quantization process is defined as:

% These embeddings are then mapped to discrete tokens using a quantizer. Specifically, the quantizer selects the closest token from a predefined codebook \( V = \{ v_i \in \mathbb{R}^D \mid i = 1, 2, \dots, K \} \), \( K \) is the number of discrete codebook vectors,
%  \( D \) is the dimension of each codebook vector. The tokenization process is described mathematically as follows: For each embedding \( \hat{h}_{j,k} \), the quantizer selects the nearest codebook vector \( v_{i^\ast} \) by minimizing the distance between the normalized embedding and the normalized codebook vectors:

\begin{equation}
\begin{aligned}
i^\ast = \mathop{\arg\min}_{i \in \{1, 2, \dotsc, K\}} \left\| \frac{h_{j,k} }{\| h_{j,k}  \|_2} - \frac{v_i}{\| v_i \|_2} \right\|^2
\end{aligned}
\end{equation}

where \( \| \cdot \|_2 \) denotes the \( \ell_2 \)-norm, and normalization ensures that the distance measure is equivalent to maximizing the cosine similarity. The assigned discrete token index for the patch \( x_{j,k} \) is index \( i^\ast \). This process effectively quantizes the ECG signal into a sequence of discrete tokens \( \{ z_{j,k} \} \), reducing the influence of noise and enhancing the signal quality.

By transforming the ECG data into a low dimension and high-SNR tokenized representations \( z_{j,k} \), the \tokenizer enables the model to focus on the meaningful aspects of the cardiac signal, such as heartbeat patterns and rhythms, which improves the model’s ability to generalize across different datasets.

% In summary, our tokenizer addresses the challenges posed by the inherent low-SNR and high-dimensional nature of ECG data, as well as the variations between datasets. By converting the raw signals into discrete, high-SNR tokens that reflect real cardiac events, we create a more robust and generalizable representation for downstream tasks such as anomaly detection and rhythm classification.

\subsubsection{Multi-View Synergistic Decoder}

To better capture the demographic variations and morphological changes inherent in ECG signals, we propose a \textbf{Multi-View Synergistic Decoder} containing three decoders for different proxy tasks.

% The decoder integrates three specialized decoders: the \textbf{Morphology Decoder}, the \textbf{Frequency Decoder}, and the \textbf{Demography Decoder}. 

% By simultaneously focusing on these aspects, our approach captures a comprehensive representation of the ECG data, enhancing the robustness and accuracy of cardiac activity analysis.

\paragraph{Morphology Decoder} aims to reconstruct the original temporal ECG signals, focusing on preserving time-domain information critical for identifying features like QRS complexes and arrhythmia. By reconstructing the time-domain signals, we ensure that the essential temporal characteristics of the cardiac cycles are retained, providing a foundation for accurate heartbeat analysis. 
The reconstruction loss for the Morphology Decoder is defined as:
\begin{equation}
\begin{aligned}
\mathcal{L}_{\text{morphology}} = \sum_{j=1}^{L} \sum_{k=1}^{P} \left\| o_{j,k}^m - x_{j,k} \right\|_2^2
\end{aligned}
\end{equation}
where \( o_{j,k}^m \) is the output of the  Morphology Decoder  for patch \( x_{j,k} \).

\paragraph{Frequency Decoder} predicts the frequency characteristics of ECG signals by incorporating frequency-domain information, which is essential for capturing periodic and spectral features associated with cardiac conditions. Unlike traditional methods that focus solely on time-domain or frequency features, this decoder leverages the Discrete Wavelet Transform (DWT)\cite{shensa1992discrete} to analyze the signals simultaneously in both time and frequency domains. For each ECG patch \( x_{j,k} \in \mathbb{R}^s \), corresponding to lead \( j \) and patch index \( k \), we apply the DWT to decompose the time-domain signal into wavelet coefficients, capturing localized frequency content. The DWT performs a multi-scale decomposition of the signal recursively, obtaining features across different frequency ranges. The wavelet decomposition process consists of two main parts. At the initial stage, the original signal is the approximation coefficients at level zero, \( c_{A}^{(0)} = x_{j,k} \). Then, at the Recursive Decomposition stage, for each level \( l \) (\( l = 1, 2, \dotsc, L_w \)), we use the approximation coefficients from the previous level \( c_{A}^{(l-1)} \) to obtain the current level's approximation coefficients \( c_{A}^{(l)} \) and detail coefficients \( c_{D}^{(l)} \) through convolution and downsampling:

% \begin{equation}
% \begin{aligned}
%   c_{A}^{(l)}[n] &= \sum_{m} c_{A}^{(l-1)}[m] \cdot g[2n - m]
% \end{aligned}
% \ \ \ \
% \begin{aligned}
%   c_{D}^{(l)}[n] &= \sum_{m} c_{A}^{(l-1)}[m] \cdot h[2n - m]
% \end{aligned}
% \end{equation}

\begin{equation}
\begin{aligned}
  c_{A}^{(l)}[n] &= \sum_{m} c_{A}^{(l-1)}[m] \cdot g[2n - m], \\[0.1em]
  c_{D}^{(l)}[n] &= \sum_{m} c_{A}^{(l-1)}[m] \cdot h[2n - m].
\end{aligned}
\end{equation}

where \( g[\cdot] \) and \( h[\cdot] \) are the coefficients of the low-pass and high-pass filters, respectively, \( n \) is the index of the downsampled coefficients, the convolution operation captures the signal's features in the corresponding frequency range, and downsampling reduces the resolution, focusing on lower-frequency components. We obtain a hybrid time-frequency representation of the ECG signal through multi-scale decomposition by performing these operations on the approximation coefficients \( c_{A}^{(l-1)} \) at each level, simultaneously capturing both the low-frequency (approximation coefficients) and high-frequency (detail coefficients) information of the signal. For stable convergence during training, we apply z-score normalization to the frequency magnitudes within each patch. The reconstruction loss for the Frequency Decoder is defined as:

\begin{equation}
\begin{aligned}
\mathcal{L}_{\text{freq}} = \sum_{l=1}^{L_w} \left( \left\| \hat{c}_{A}^{(l)} - c_{A}^{(l)\, \text{norm}} \right\|_2^2 + \left\| \hat{c}_{D}^{(l)} - c_{D}^{(l)\, \text{norm}} \right\|_2^2 \right)
\end{aligned}
\end{equation}

where \( \hat{c}_{A}^{(l)} \) and \( \hat{c}_{D}^{(l)} \) are the predicted approximation and detail coefficients at level \( l \), respectively, and \( c_{A}^{(l)\, \text{norm}} \) and \( c_{D}^{(l)\, \text{norm}} \) are the corresponding normalized actual coefficients. The loss is computed across all decomposition levels \( l \) from 1 to \( L_w \).

\paragraph{Demography Decoder} predicts patient-specific attributes (e.g., age, weight, or other demographic factors), represented as a vector \( a \in \mathbb{R}^{d_a} \). By jointly predicting these attributes, the model gains a personalized understanding of the patient's condition.  This personalized aspect allows the model to better account for inter-patient variability, which is critical in making accurate clinical predictions. The loss for the Demography Decoder is defined as:

\begin{equation}
\begin{aligned}
\mathcal{L}_{\text{demography}} = \left\| o^a - a \right\|_2^2
\end{aligned}
\end{equation}

where \( o^a \) represents the predicted patient-specific attributes, and \( a \) is the ground truth patient attribute vector.

\paragraph{Overall Loss Function for \tokenizer} In addition to reconstruction loss functions from all decoders, we also include codebook loss and commitment loss to ensure that the quantized tokens remain faithful to the original signal and stabilize the training process. The codebook loss and commitment loss are defined as:

% \begin{equation}
% \begin{aligned}
% \mathcal{L}_{\text{codebook}} = \sum_{j=1}^{C} \sum_{k=1}^{P} \left\| \operatorname{sg} \left( h_{j,k} \right) - v_{z_{j,k}} \right\|_2^2
% \end{aligned}

% \begin{aligned}
% \mathcal{L}_{\text{commitment}} = \beta \sum_{j=1}^{C} \sum_{k=1}^{P} \left\| h_{j,k} - \operatorname{sg} \left( v_{z_{j,k}} \right) \right\|_2^2
% \end{aligned}
% \end{equation}
\begin{equation}
\begin{aligned}
\mathcal{L}_{\text{codebook}} &= \sum_{j=1}^{C} \sum_{k=1}^{P} \left\| \operatorname{sg} \left( h_{j,k} \right) - v_{z_{j,k}} \right\|_2^2, \\[0.1em]
\mathcal{L}_{\text{commitment}} &= \beta \sum_{j=1}^{C} \sum_{k=1}^{P} \left\| h_{j,k} - \operatorname{sg} \left( v_{z_{j,k}} \right) \right\|_2^2.
\end{aligned}
\end{equation}

where \( h_{j,k} \) is the embedding of the patch \( x_{j,k} \), \( v_{z_{j,k}} \) is the codebook vector corresponding to the token \( z_{j,k} \), \( \operatorname{sg}(\cdot) \) denotes the stop-gradient operator, and \( \beta \) is a weighting coefficient for the commitment loss. The overall loss function combines all components:

% \begin{equation}
% \begin{aligned}
% \mathcal{L}_T = \mathcal{L}_{\text{morphology}} + \mathcal{L}_{\text{frequency}} + \mathcal{L}_{\text{demography}} + \mathcal{L}_{\text{codebook}} + \mathcal{L}_{\text{commitment}}
% \end{aligned}
% \end{equation}
\begin{equation}
\begin{split}
\mathcal{L}_T = &\ \mathcal{L}_{\text{morphology}} + \mathcal{L}_{\text{frequency}} + \mathcal{L}_{\text{demography}} + \\ &\mathcal{L}_{\text{codebook}} + \mathcal{L}_{\text{commitment}}
\end{split}
\end{equation}

This loss function requires the reconstruction of both the temporal and frequency components of the ECG signal, while also ensuring the recovery of patient-specific factors for personalized modeling. Experiments in Appendix~\ref{sec:ablation_loss} shows the importance of each component in the total loss function.

\subsection{\model Masked Pre-training}

% \subsubsection{Masked ECG Modeling}

Inspired by self-supervised learning from masked modeling in NLP~\cite{kenton2019bert} and vision~\cite{bao2021beit,he2022masked}, we design a hybrid-scale masked ECG modeling strategy, where random segments of ECG signals are masked and the model is learned to reconstruct missing parts. 

% Our method builds upon masked ECG modeling and multi-channel autoregressive strategies, combining the strengths of both techniques to better capture temporal and spatial dependencies across ECG leads.

% After using \tokenizer process $X$ to embeddings \( h_{j,k} \in \mathbb{R}^d \), we apply a random masking procedure. For each embedding, a mask \( m_{j,k} \in \{0, 1\} \) is generated such that a proportion \( r \) of the patches are masked (\( m_{j,k} = 1 \)). The masked patches are replaced with a learnable mask token \( h_M \in \mathbb{R}^d \). The masked embeddings \( \tilde{h}_{j,k} \) are defined as: $\tilde{h}_{j,k} = (1 - m_{j,k}) \cdot h_{j,k} + m_{j,k} \cdot h_M$. To incorporate positional information, we augment the masked embeddings with temporal and spatial positional embeddings: $\bar{h}_{j,k} = \tilde{h}_{j,k} + \tau_k + \sigma_j$. These augmented embeddings \( \tilde{h}_{j,k} \) are then reshaped into a sequential format and fed into a Transformer encoder to generate contextualized representations \( \tilde{h}'_{j,k} \in \mathbb{R}^d \). Each contextualized vector \( \tilde{h}'_{j,k} \) is passed through a linear classifier followed by a softmax function to produce a probability distribution over the codebook tokens \( V = \{ v_1, v_2, \dotsc, v_K \} \):

After using \tokenizer process $X'$ to embeddings $H \in \mathbb{R}^{N \times d}$, we randomly generate a mask $M \in \mathbb{R}^{N \times 1}$, where its component \( m_{j,k} \in \{0, 1\} \). The masked patches are replaced with a learnable mask token \( h_M \in \mathbb{R}^d \). The masked embeddings \( \tilde{h}_{j,k} \) are defined as: $\tilde{h}_{j,k} = (1 - m_{j,k}) \cdot h_{j,k} + m_{j,k} \cdot h_M$. These augmented embeddings \( \tilde{h}_{j,k} \) are then reshaped into a sequential format and fed into a Transformer encoder to generate contextualized representations \( \tilde{h}'_{j,k} \in \mathbb{R}^d \). Each contextualized vector \( \tilde{h}'_{j,k} \) is passed through a linear classifier followed by a softmax function to produce a probability distribution over the codebook tokens \( V = \{ v_1, v_2, \dotsc, v_K \} \): $p(v_i \mid \tilde{H}) = \text{softmax}\left( W \bar{h}'_{j,k} + b \right)_i $, where \( \bar{H} \) denotes the collection of all augmented embeddings \( \tilde{h}_{j,k} \), and the subscript \( i \) refers to the \( i \)-th element of the output vector. The training objective for the masked modeling process is to minimize the negative log-likelihood of predicting the correct tokens \( v_{z_{j,k}} \) at the masked positions:

\begin{equation}
\begin{aligned}
\mathcal{L}_\text{mask} = - \sum_{j=1}^{L} \sum_{k=1}^{P} m_{j,k} \cdot \log p\left( v_{z_{j,k}} \mid \tilde{H} \right)
\end{aligned}
\end{equation}

The masked pre-training facilitates the model in learning generic representations from the input data by capturing the implicit rhythm-event associations and sequential relationships crucial for ECG analysis,  thereby enhancing its ability to capture the underlying cardiac event patterns in the ECG signals. 

% By employing this approach, the model becomes more adept at predicting accurate high-level features, allowing it to discern intricate rhythm patterns and event-level semantics, thereby enhancing its ability to capture the underlying cardiac event patterns in the ECG signals. 

% \subsubsection{Linear Attention Encoder}

% Finally, we use a linear attention encoder to reduce the computational complexity of the model. This encoder is designed to handle long sequences efficiently, making it suitable for real-time ECG analysis in clinical settings. The linear attention mechanism helps the model focus on critical signal patterns while discarding irrelevant noise.

\subsection{Takeaway}
Our proposed \model framework innovatively addresses three core challenges in current ECG Foundation Models, with the following key insights:

% \textbf{Heterogeneity} in ECG signals—arising from varying sampling rates, lead configurations, and recording lengths—is addressed through standardized preprocessing(to unify signal resolution) and CSA(to tackle computational inefficiency in ultra-long ECG). Input signals are resampled to 300 Hz and segmented into fixed-duration patches, ensuring compatibility across devices. \textcolor{red}{CSA resolves this by selectively attending to critical rhythm phases and cross-lead dependencies, mimicking clinicians’ diagnostic focus.  This sparse attention mechanism reduces redundant computation on non-informative segments, enabling efficient analysis of ultra-long sequences without sacrificing temporal resolution—a key limitation in prior work.}

\textbf{Heterogeneity} in ECG signals—arising from varying sam-pling rates, lead configurations, and recording lengths—poses significant challenges for consistent analysis. To address this, standardized preprocessing is employed to unify signal resolution, ensuring compatibility across diverse devices by resampling signals to 300 Hz and segmenting them into fixed-duration patches.  However, this preprocessing step introduces a side effect: the generation of long ECG patches, which can lead to computational inefficiency.  To resolve this issue, CSA is introduced, which selectively focuses on critical rhythm phases and cross-lead dependencies, mimicking the diagnostic focus of clinicians.  This sparse attention mechanism reduces redundant computations on non-informative segments, enabling efficient analysis of ultra-long sequences without sacrificing temporal resolution—a key limitation in prior work.  

\textbf{Low SNR}, which obscures subtle pathological features, is mitigated by the Rhythm Quantizer. This module distills ECG signals into discrete tokens through three reconstruction objectives: (1) morphological reconstruction of disease-specific waveforms, (2) time-frequency alignment via wavelet decomposition to isolate rhythm irregularities (3) demographic attribute recovery (age-related T-wave variations). By explicitly encoding  cardiac rhythms, the quantizer disentangles clinically actionable patterns from interference.

\textbf{Demographic shifts}—such as age, sex, and ethnicity biases—are tackled via a two-stage pre-training paradigm. First, local rhythm codes are learned to capture demographic-agnostic patterns. Second, masked modeling of global cardiac semantics links rhythm events to clinical contexts, ensuring robustness across populations. The model decouples age-related variations from pathological signatures. This approach bridges domain gaps between hospital-grade and wearable ECG data.

CSA and Rhythm Quantizer jointly address heterogeneity and noise, while two-stage pre-training ensures demographic invariance. The unified design establishes a foundation for scalable, device-agnostic cardiac diagnostics.

%% file: sec/5.experiment.tex
\section{Downstream Application}

\begin{figure*}[t]
    \centering
    \includegraphics[width=0.99\textwidth]{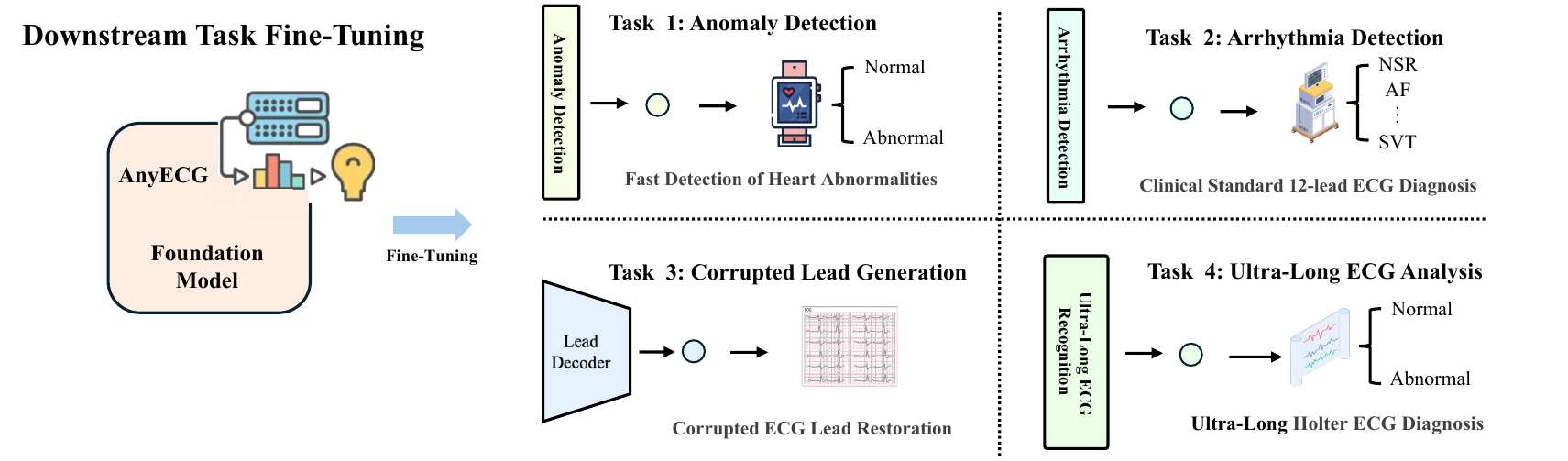}
    \caption{\textbf{Downstream Task of the ECG Foundation Model.} 
 The framework demonstrates fine-tuning \model on heterogeneous ECG datasets, including portable devices, Holter monitors, standard 12-lead recordings, and Event monitor, to address four critical tasks: (1) Anomaly Detection (binary classification of normal vs. abnormal rhythms), (2) Arrhythmia Classification, (3) Ultra-Long ECG Analysis (continuous monitoring for ultra-long ECG event detection), and (4) Corrupted Lead Generation. \model achieves task-specific adaptability through fine-tuning while preserving pre-learned cardiac semantics.
    % Additionally, a complementary masking approach was proposed to provide more masking perspectives focusing on both heavily masked and minimally masked regions of the ECG signals.
    }
    \label{fig:downstream}
\end{figure*}

This section evaluates \model's performance across multiple ECG datasets to prove its generality. As illustrated in Figure~\ref{fig:downstream}, \model is fine-tuned on heterogeneous ECG data to address four critical tasks: anomaly detection, arrhythmia classification, corrupted lead generation, and ultra-long ECG recognition. In Section~\ref{subsec:dataset}, we summarize datasets utilized in the experiments. Section~\ref{subsec:exp_setup} explains the experimental setup in detail. In Section~\ref{subsec:result}, we present the results of our experiments, benchmarking \model against state-of-the-art methods across multiple tasks, including anomaly detection, arrhythmia detection, ECG lead generation, and ultra-long ECG sequence recognition. In Section~\ref{sec:ablation_loss} and \ref{sec:ablation_pretrain}, we also present \textbf{ablation studies} on hyperparameter selection and the necessity of two-stage pre-training.

\subsection{ECG Datasets}
\label{subsec:dataset}

To evaluate the performance of \model and baseline models, we utilized a comprehensive set of ECG datasets that include all available unlabeled data during pre-training. These datasets cover a wide spectrum of cardiac conditions, patient demographics, and recording scenarios, ensuring robust testing across diverse settings. For various downstream tasks, we mixed all datasets together to minimize biases introduced by individual datasets and to better validate the model's generalizability. This approach reduces the discrepancies arising from different data sources and enhances the unified capability of the model. The detailed data construction of the datasets can be found in Table~\ref{tab:ecg_datasets}. All datasets are formatted in WFDB format, including associated binary and text files that detail signal attributes and clinical annotations using SNOMED-CT codes.

To evaluate the performance of \model and baseline models, we utilized a comprehensive set of ECG datasets that cover a wide spectrum of cardiac conditions, patient demographics, and recording scenarios, ensuring robust testing across diverse settings. The datasets include:

\textbf{CPSC and CPSC-Extra Databases~\cite{liu2018open}:} These consist of 12-lead ECG recordings ranging from 6 to 60 seconds in duration, sampled at 500 Hz, and include a balanced mix of male and female subjects.

\textbf{INCART Database~\cite{tihonenko2008st}:} This database provides 74 annotated recordings extracted from 32 Holter records, each 30 minutes long and sampled at 257 Hz, offering high-resolution data ideal for arrhythmia classification.

\textbf{PTB~\cite{bousseljot1995nutzung} and PTB-XL Databases~\cite{wagner2020ptb}:} The PTB Diagnostic ECG Database includes 516 recordings sampled at 1000 Hz, while PTB-XL contains 21,837 12-lead ECGs sampled at 500 Hz, each 10 seconds long, encompassing a wide range of cardiac pathologies.

\textbf{Georgia 12-lead ECG Challenge (G12EC) Database:} Comprising 10,344 recordings from the Southeastern United States, sampled at 500 Hz, this dataset adds demographic diversity to our evaluation.

\textbf{Undisclosed Database:} This dataset contributes an additional 10,000 ECG recordings, providing a geographically distinct test set to further validate the model's performance without data leakage.

% All datasets are formatted in WFDB format, including associated binary and text files that detail signal attributes and clinical annotations using SNOMED-CT codes.

By employing this diverse collection of datasets, we thoroughly assess \model's ability to generalize across different patient populations, signal qualities, and clinical conditions.

% \subsection{Evaluation Metrics.}\label{sec:Metrics}
% We comprehensively evaluated our model using set of metrics, including cardiac disease classification \textbf{accuracy}, \textbf{precision}: The proportion of true positive predictions out of all positive predictions, indicating the model's ability to avoid false positives. \textbf{AUC-PR}: The Area Under the Precision-Recall curve, useful for binary classification with imbalanced data. \textbf{AUROC}: The Area Under the Receiver Operating Characteristic curve, reflecting the model's ability to distinguish between classes. \textbf{Weighted F1 Score}: The harmonic mean of precision and recall, weighted by class size. \textbf{PSNR}: The Peak Signal-to-Noise Ratio, which measures the quality of signal reconstruction, with higher values indicating better quality. \textbf{SSIM}: The Structural Similarity Index, which assesses structural similarity between original and reconstructed signals, with values closer to 1 indicating higher similarity. \textbf{MAE}: The Mean Absolute Error, representing the average absolute difference between predicted and actual values, indicating prediction accuracy.

\subsection{Evaluation Metrics}\label{sec:Metrics}

We conducted four distinct experiments, each utilizing a specific set of evaluation metrics tailored to the task:

1.  \textbf{Anomaly Detection:} Evaluated using \textbf{Accuracy}, \textbf{AUC-PR} (Area Under the Precision-Recall Curve), \textbf{AUROC} (Area Under the Receiver Operating Characteristic Curve), and \textbf{Weighted F1 Score}. These metrics assess the model’s ability to correctly identify anomalies and handle class imbalances effectively.

2.  \textbf{Arrhythmia Detection:} Assessed with \textbf{Accuracy}, \textbf{AUC-PR}, \textbf{Weighted F1 Score}, and \textbf{Precision}. This combination of metrics evaluates the model’s performance in detecting various types of arrhythmias, emphasizing both overall accuracy and the precision of positive predictions.

3.  \textbf{Corrupted Lead Generation:} Measured using \textbf{PSNR} (Peak Signal-to-Noise Ratio), \textbf{SSIM} (Structural Similarity Index), and \textbf{MAE} (Mean Absolute Error). These metrics quantify the quality of the generated ECG signals by comparing the reconstructed signals to the original ones, focusing on signal fidelity and structural similarity.

4.  \textbf{Ultra-Long ECG Analysis:} Evaluated using \textbf{Accuracy}, \textbf{AUC-PR}, \textbf{AUROC}, and \textbf{Weighted F1 Score}. These metrics measure the model's effectiveness in accurately identifying anomalies and handling class imbalances.

The definitions of the evaluation metrics used across these experiments are as follows: \textbf{Accuracy}: The proportion of correctly predicted instances out of all instances, indicating the overall effectiveness of the model.
 \textbf{Precision}: The ratio of true positive predictions to the total number of positive predictions, reflecting the model's ability to avoid false positives.
 \textbf{AUC-PR} (Area Under the Precision-Recall Curve): Measures the trade-off between precision and recall for different threshold settings, particularly useful for imbalanced datasets.
 \textbf{AUROC} (Area Under the Receiver Operating Characteristic Curve): Represents the model’s ability to distinguish between classes across all classification thresholds.
 \textbf{Weighted F1 Score}: The harmonic mean of precision and recall, weighted by the number of true instances for each class, providing a balanced evaluation of the model’s performance.
\textbf{PSNR} (Peak Signal-to-Noise Ratio): Indicates the quality of signal reconstruction by comparing the maximum possible signal power to the power of reconstruction noise, with higher values signifying better quality.
\textbf{SSIM} (Structural Similarity Index): Assesses the similarity between two signals in terms of luminance, contrast, and structure, with values closer to 1 indicating higher similarity.
\textbf{MAE} (Mean Absolute Error): Represents the average absolute difference between predicted and actual values, serving as a measure of prediction accuracy. By employing these tailored metrics across different experiments, we ensure a comprehensive evaluation of our model’s performance in various aspects of ECG signal processing and classification.

% \paragraph{Pre-training Datasets}

% All datasets are formatted in WFDB format, including associated binary and text files that detail signal attributes and clinical annotations using SNOMED-CT codes. By employing this diverse collection of datasets, we thoroughly assess \model's ability to generalize across different patient populations, signal qualities, and clinical conditions.

% {\color{red}{TODO}}

% \paragraph{Downstream Tasks} 
% {\color{red}{TODO}}

\begin{table*}[!t]
    \centering
    \caption{Summary of ECG Datasets}
    \label{tab:ecg_datasets}
    \renewcommand{\arraystretch}{1.2} % 调整行高
    \adjustbox{max width=\textwidth}{
    \begin{tabular}{lcccc}
        \toprule[1.5pt]
        \textbf{Dataset} & \textbf{Recordings} & \textbf{Sampling Rate} & \textbf{Duration} & \textbf{Notes} \\
        \hline
        CPSC~\cite{liu2018open} & 6877 & 500 Hz  & 6--60 s & Balanced sex \\
        % \hline
        CPSC-Extra~\cite{liu2018open} & 3453 & 500 Hz  & 6--60 s & Balanced sex \\
        % \hline
        INCART~\cite{tihonenko2008st} & 74 & 257 Hz  & 30 mins & High-res; arrhythmia \\
        % \hline
        PTB~\cite{bousseljot1995nutzung} & 516 & 1000 Hz  & Varies & Wide range of pathologies \\
        % \hline
        PTB-XL~\cite{wagner2020ptb} & 21837 & 500 Hz  & 10 s & Extensive clinical ECGs \\
        % \hline
        G12EC & 10344 & 500 Hz  & Varies & The Southeast's unique demographics\\
        % \hline
        Undisclosed Dataset & 10000 &500 Hz & 6--60 s & Geographically distinct test set \\
        \bottomrule[1.5pt]
    \end{tabular}
    }
\vskip -10pt
\end{table*}

\subsection{Experimental Setup}
\label{subsec:exp_setup}

\paragraph{Model Configurations} 
We introduce three configurations of \model: \model-B, \model-L, and \model-XL, containing 254M, 500M, and 1.7B parameters, respectively. The increase in parameters is achieved by deepening the Transformer encoder and expanding the hidden layer sizes. 
% Detailed architectural settings are provided in the code. 
To maintain consistency across all configurations, we set the patch size \( P=300 \), which corresponds to 1 second of ECG data. The maximum sequence length is fixed at 1,024 tokens, sufficient for most ECG applications. During \tokenizer training and \model pre-training, sequences shorter than this length are padded. To preserve data integrity, we mask the attention values associated with these padding tokens.

% \paragraph{ECG Data Preprocessing.} To preserve the natural characteristics of the ECG signals, we applied minimal preprocessing steps.  Initially, we filtered the signals using a bandpass filter between 0.1 Hz and 75 Hz to remove low-frequency noise and a notch filter at 50 Hz to eliminate power-line interference. Subsequently, all signals were resampled to 300 Hz to ensure uniformity across the dataset. To further reduce noise, we employed a wavelet transform-based denoising method using the `db6' wavelet. In this process, the high-frequency components and the lowest approximation coefficients were set to zero to minimize noise artifacts.  Finally, the signals were reconstructed using the inverse wavelet transform.  This approach effectively reduces noise while preserving the essential features of the ECG signals.

\paragraph{Training Environment} The pre-training of \model was conducted on a comprehensive dataset compiled from seven different sources. For the downstream tasks, data splitting followed standard procedures, dividing the data into training and validation subsets using an 80/20 ratio. Binary cross-entropy loss was employed for binary classification tasks, while cross-entropy loss was utilized for multi-class classification tasks. 
% Evaluation metrics for the downstream tasks are detailed in the Appendix \ref{sec:Metrics}. 
All experiments were executed on a computing cluster equipped with eight high-performance GPUs. We used the Adam optimizer with a learning rate of 1e-4 for all models training. Model selection was based on the best performance on validation sets, and final evaluations were conducted on separate test sets. To ensure the reliability of our results, performance metrics—including averages and standard deviations—were reported across five random seeds.

\subsection{Experimental Results}
\label{subsec:result}

\paragraph{Anomaly Detection}
In anomaly detection, we define normal category data as positive samples, with all other categories treated as anomalous samples. Table~\ref{tab:ecg_anomaly} compares \model to state-of-the-art models in the anomaly detection task. \model consistently outperforms other advanced models across all evaluation metrics. Specifically, the largest variant, \model-XL, achieves the highest scores in accuracy, AUC-PR, AUROC, and Weighted F1 Score, demonstrating its strong ability to capture ECG signal characteristics. In contrast, traditional models like DENS-ECG~\cite{peimankar2021dens} and ContraWR~\cite{yang2021self} show lower performance. DENS-ECG achieves moderate scores in accuracy and Weighted F1 Score, while ContraWR falls short in both metrics. Even the smaller versions of \model, such as \model-B and \model-L, perform competitively and surpass most baseline models. This indicates that \model maintains high performance across different scales without requiring extensive model parameters. Notably, the finetuned ECG-FM model~\cite{mckeen2024ecg} performs at an intermediate to above-average level compared to the baseline. However, as a pre-trained model, its performance may still be hindered by substantial differences between the pre-training data and the downstream task dataset, which likely impedes its ability to fully converge.

% Table \ref{tab:ecg_anomaly} provides a comprehensive comparison of \model with state-of-the-art models in the anomaly detection task. As shown, \model consistently demonstrates superior performance across all evaluation metrics, outperforming other advanced models. Specifically, the largest variant, \textbf{\model-XL}, achieves the highest scores in accuracy, AUC-PR, AUROC, and Weighted F1 Score, indicating \model's strong capability in capturing ECG signal characteristics. In contrast, traditional models like DENS-ECG and ContraWR show lower performance, with DENS-ECG achieving moderate scores in accuracy and Weighted F1 Score, while ContraWR also falls short in both metrics. Even the smaller versions of \model, such as \model-B and \model-L, demonstrate competitive performance, surpassing most baseline models. For example, \model-B achieves a considerably higher Weighted F1 Score compared to models like ST-Transformer and Inception1D, emphasizing that \model maintains high performance across different scales without requiring extensive model parameters. Notably, the pre-trained ECG-FM model shows the lowest performance among all models in both accuracy and Weighted F1 Score. This is likely due to the significant distributional differences between the pre-training data and the downstream task dataset, making it difficult for the model to converge. This contrast further underscores the effectiveness of \model's pre-training approach in capturing generalized ECG features.

\begin{table*}[h]
\centering
\caption{Results Comparison with State-of-the-Art Models in Anomaly Detection}
\label{tab:ecg_anomaly}
\adjustbox{max width=\textwidth}{
\begin{tabular}{cccccc}
\toprule
\textbf{Methods} &\textbf{Pretrain}& \textbf{Accuracy} $\uparrow$ & \textbf{AUC-PR} $\uparrow$ & \textbf{AUROC} $\uparrow$ & \textbf{Weighted F1 Score} $\uparrow$ \\
\midrule
DENS-ECG~\cite{peimankar2021dens} & \xmark & 0.7928±0.0019 & 0.9319±0.0019 & 0.8488±0.0070 & 0.7928±0.0009 \\
ContraWR~\cite{yang2021self} & \xmark &0.7551±0.0011 & 0.9374±0.0001 & 0.8153±0.0002 & 0.7611±0.0003 \\
XResNet1D~\cite{he2019bag} & \xmark &0.7768±0.0115 & 0.9217±0.0045 & 0.7522±0.0121 & 0.7606±0.0093 \\
CNN-Transformer~\cite{peh2022transformer} & \xmark & 0.7401±0.0019 & 0.9340±0.0011 & 0.8074±0.0034 & 0.7444±0.0005 \\
RNN1D~\cite{salloum2017ecg} & \xmark &0.7992±0.0017 & 0.9284±0.0006 & 0.7868±0.0015 & 0.7838±0.0012 \\
FFCL~\cite{li2022motor} & \xmark &0.6709±0.0012 & 0.8682±0.0003 & 0.6423±0.0018 & 0.6746±0.0003 \\
Inception1D~\cite{strodthoff2020deep} & \xmark &0.8001±0.0029 & 0.9408±0.0004 & 0.8097±0.0015 & 0.7868±0.0018 \\
ST-Transformer~\cite{song2021transformer} & \xmark & 0.8070±0.0017 & 0.9471±0.0007 & 0.8406±0.0004 & 0.8048±0.0004 \\
ECG-FM~\cite{mckeen2024ecg} & \cmark &0.7788±0.0029 & 0.9036±0.0197 & 0.7693±0.0028 & 0.7321±0.0112 \\
\midrule
\model-B & \cmark &0.8188±0.0025 & 0.9517± 0.0049 & 0.8502±0.0026 & 0.8863±0.0022 \\
\model-L & \cmark &0.8241±0.0043 & 0.9535±0.0030 & 0.8483±0.0025 & 0.8898±0.0026 \\
\model-XL & \cmark &\textbf{0.8255±0.0035} & \textbf{0.9538±0.0012} & \textbf{0.8550±0.0016} & \textbf{0.8912±0.0033} \\
\bottomrule
\end{tabular}
}
\vskip -10pt
\end{table*}

\begin{table*}[h]
\centering
\caption{Results Comparison with State-of-the-Art Models in Ultra-Long ECG Analysis}
\label{tab:long}
\adjustbox{max width=\textwidth}{
\begin{tabular}{cccccc}
\toprule
\textbf{Methods} & \textbf{Adaptation} &\textbf{Accuracy} $\uparrow$ & \textbf{AUC-PR} $\uparrow$ & \textbf{AUROC} $\uparrow$ & \textbf{Weighted F1 Score} $\uparrow$ \\
\midrule
DENS-ECG~\cite{peimankar2021dens} & \xmark &0.3202±0.0074 & 0.1514±0.0042 & 0.2669±0.0085 & 0.2866±0.0069 \\
ContraWR~\cite{yang2021self} & \xmark & 0.3075±0.0035 & 0.1359±0.0048 & 0.2802±0.0055 & 0.2794±0.0083 \\
XResNet1D~\cite{he2019bag} & \xmark &0.6611±0.0812 & 0.6916±0.0797 & 0.6499±0.1353 & 0.6453±0.0922 \\
CNN-Transformer~\cite{peh2022transformer} & \xmark & 0.3284±0.0202 & 0.1417±0.0071 & 0.2685±0.0290 & 0.2641±0.0061 \\
RNN1D~\cite{salloum2017ecg} & \xmark &0.7444±0.0102 & 0.7724±0.0102 &0.8679±0.0291 & 0.7386±0.0640 \\
FFCL~\cite{li2022motor} & \xmark & 0.1823±0.0035 & 0.0832±0.0050 & 0.1770±0.0052 & 0.1736±0.0013 \\
Inception1D~\cite{strodthoff2020deep} & \xmark &0.5000±0.0017 & 0.5154±0.0492 & 0.3197±0.0573 & 0.3432±0.0038 \\
ST-Transformer~\cite{song2021transformer} & \xmark & 0.2011±0.0057 & 0.0941±0.0046 & 0.1996±0.0053 & 0.2018±0.0027 \\
% ECG-FM~\cite{mckeen2024ecg} & \xmark & - & - & - & - \\
\midrule
\model-B & \cmark & 0.6944±0.0016 & 0.7482±0.0025 & 0.6759±0.0056 & 0.5639±0.0124 \\
\model-L & \cmark & 0.7777±0.0077 & 0.9075±0.0072 & 0.9104±0.0039 & 0.7500±0.0072 \\
\model-XL & \cmark & \textbf{0.8055±0.0034} & \textbf{ 0.9088±0.0027} & \textbf{0.9104±0.0147} & \textbf{0.7741±0.0068} \\
\bottomrule
\end{tabular}
}
\vskip -10pt
\end{table*}

\paragraph{Arrhythmia Detection}
Table \ref{tab:arrhythmia} presents a performance comparison between \model and other leading models in arrhythmia detection. The results show that \model, particularly the \model-XL variant, consistently outperforms competing models across all metrics. This demonstrates its strong ability to handle arrhythmia detection effectively. In contrast, models like DENS-ECG~\cite{peimankar2021dens} and ContraWR~\cite{yang2021self} exhibit lower performance. Notably, although ECG-FM~\cite{mckeen2024ecg} employs pre-training, it achieves significantly lower accuracy. 
% The absence of results for the ECG-FM model~\cite{mckeen2024ecg}, indicated by "-", is due to its inability to converge on this 70-class label task with severe class imbalance. 
This underscores \model's robustness, as its consistent performance across all metrics confirms its suitability for real-world arrhythmia detection.

% To evaluate the effectiveness of \model in arrhythmia detection, we compare its performance against other state-of-the-art models. Tbale \ref{tab:arrhythmia} shows the performance comparison of \model with other leading models in the arrhythmia detection task. The results demonstrate that \model, particularly the \textbf{\model-XL} variant, consistently outperforms competing models across all metrics, indicating its strong ability to handle arrhythmia detection effectively. Models like DENS-ECG and ContraWR exhibit relatively lower performance, while even the smaller versions of \model, such as \model-B and \model-L, outperform most baseline models, highlighting \model's adaptability across different scales. The absence of ECG-FM results, marked by "-", is due to the model's inability to converge in this 70-class label task with severe class imbalance. In contrast, \model's consistent performance across all metrics reaffirms its robustness and suitability for real-world arrhythmia detection applications.

\begin{table*}[h]
\centering
\caption{Results Comparison with State-of-the-Art Models in Arrhythmia Detection}
\label{tab:arrhythmia}
\adjustbox{max width=\textwidth}{
\begin{tabular}{cccccc}
\toprule
\textbf{Methods} & \textbf{Pretrain} &\textbf{Accuracy} $\uparrow$ & \textbf{AUC-PR} $\uparrow$ & \textbf{Weighted F1 Score} $\uparrow$ & \textbf{Precision} $\uparrow$ \\
\midrule
DENS-ECG~\cite{peimankar2021dens} & \xmark &0.3202±0.0074 & 0.1514±0.0042 & 0.2669±0.0085 & 0.2866±0.0069 \\
ContraWR~\cite{yang2021self} & \xmark & 0.3075±0.0035 & 0.1359±0.0048 & 0.2802±0.0055 & 0.2794±0.0083 \\
XResNet1D~\cite{he2019bag} & \xmark &0.1822±0.0058 & 0.1044±0.0011 & 0.1765±0.0031 & 0.1746±0.0124 \\
CNN-Transformer~\cite{peh2022transformer} & \xmark & 0.3284±0.0202 & 0.1417±0.0071 & 0.2685±0.0290 & 0.2641±0.0061 \\
RNN1D~\cite{salloum2017ecg} & \xmark &0.2511±0.0019 & 0.0911±0.0005 & 0.2164±0.0011 & 0.1986±0.0010 \\
FFCL~\cite{li2022motor} & \xmark & 0.1823±0.0035 & 0.0832±0.0050 & 0.1770±0.0052 & 0.1736±0.0013 \\
Inception1D~\cite{strodthoff2020deep} & \xmark &0.2770±0.0031 & 0.1280±0.0006 & 0.2487±0.0031 & 0.2371±0.0021 \\
ST-Transformer~\cite{song2021transformer} & \xmark & 0.2011±0.0057 & 0.0941±0.0046 & 0.1996±0.0053 & 0.2018±0.0027 \\
ECG-FM~\cite{mckeen2024ecg} & \cmark & 0.2212±0.0015 & 0.1037±0.0042 & 0.2285±0.0064 & 0.2386±0.0153\\
\midrule
\model-B & \cmark & 0.3339±0.0029 & 0.1524±0.0069 & 0.2747±0.0046 & 0.3350±0.0052 \\
\model-L & \cmark & 0.3358±0.0077 & 0.1542±0.0035 & 0.2636±0.0040 & 0.3339±0.0080 \\
\model-XL & \cmark & \textbf{0.3449±0.0095} & \textbf{0.1635±0.0028} & \textbf{0.2833±0.0033} & \textbf{0.3449±0.0075} \\
\bottomrule
\end{tabular}
}
\vskip -10pt
\end{table*}

\paragraph{Corrupted Lead Generation}

\begin{figure*}[t]
    \centering
    \includegraphics[width=0.99\textwidth]{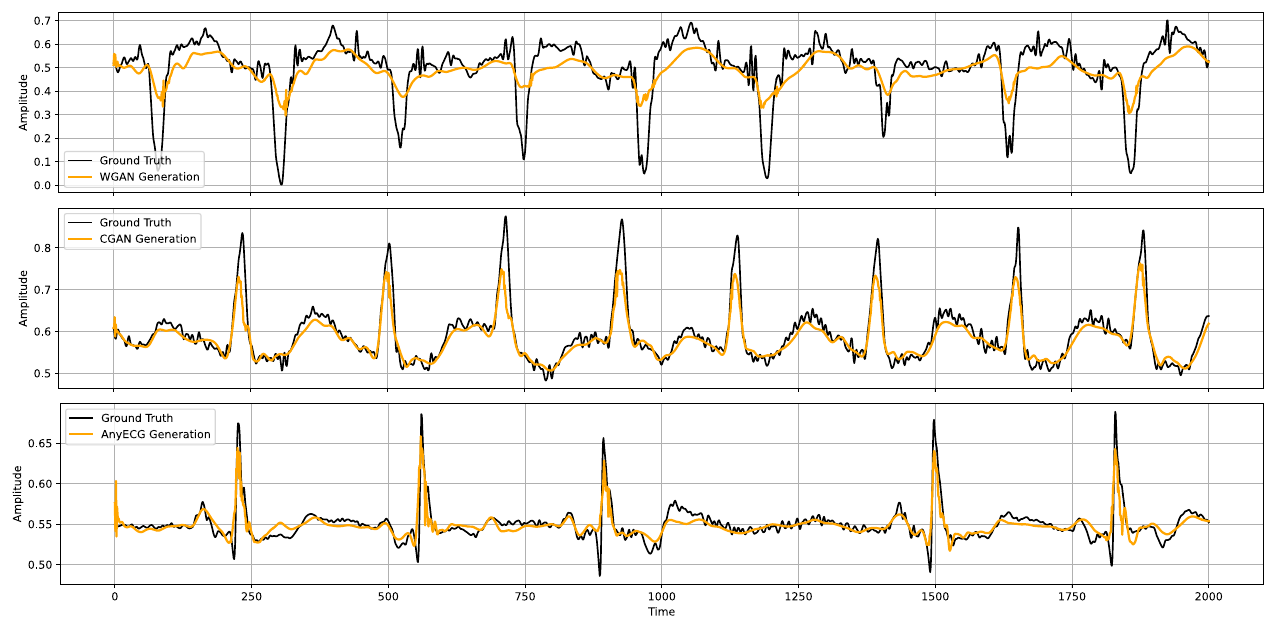}
    \caption{\textbf{Visualization of Corrupted Lead Generation} among WGAN (top), CGAN (middle), {\model} (bottom).}
    \label{fig:generation}
    \vskip -10pt
\end{figure*}

We evaluated \model against CGAN~\cite{mirza2014conditional} and WGAN~\cite{adler2018banach} in generating corrupted ECG leads (see Table~\ref{tab:Generation} and Figure~\ref{fig:generation}). Using metrics like PSNR, SSIM, and MAE, \model-L achieved the highest PSNR (32.7372 dB) and SSIM (0.8738), outperforming both CGAN and WGAN. Smaller models like AnyECG-L and AnyECG-B offer a better balance between capacity and generalization compared to AnyECG-XL. Due to limitations in its model architecture, ECG-FM~\cite{mckeen2024ecg} could not be applied to this task. Although the AnyECG models did not achieve the lowest MAE, this may be because they prioritize capturing abstract rhythms and morphological patterns over minimizing pixel-level errors in detailed, noisy signals. This suggests that while AnyECG effectively captures the overall structure and rhythm of ECG signals, it is somewhat less precise in reproducing finer details. Figure~\ref{fig:generation} shows the ECG signals generated by WGAN, CGAN, and AnyECG. Both WGAN and CGAN can capture general morphology but fail to accurately reproduce certain rhythms, leading to unsuccessful signal generation in those cases. AnyECG leverages two stage pre-training to capture complex rhythmic features, resulting in morphology closer to the original signals. However, it lacks detailed feature extraction in finer wave bands, leading to poorer reconstruction in these regions and higher MAE. These observations suggest that while AnyECG excels in preserving overall rhythmic and morphological integrity, there is room for improvement in reconstructing fine-grained details.

% The larger \model-XL did not perform best, possibly due to overfitting or excessive complexity; 
% Compared to \model-XL, smaller models like \model-L and \model-B offer a better balance between capacity and generalization. 
% Although \model did not achieve the lowest MAE, this may be because it focuses more on abstract rhythms and morphological representations rather than reducing pixel-wise errors in detailed signals that include noise. This indicates that while \model is effective in capturing the overall structure and rhythm of ECG signals, it is slightly less precise in details. The results confirm \model's effectiveness in efficiently generating high-quality ECG signal reconstructions. Figure~\ref{fig:generation} shows the ECG signals generated by WGAN, CGAN, and AnyECG. Both WGAN and CGAN can capture general morphology but fail to accurately reproduce certain rhythms where signal generation was unsuccessful. AnyECG leverages pre-training to capture high-dimensional rhythmic features, resulting in morphology closer to the original signals. However, it lacks detailed feature extraction in finer wave bands, leading to poorer reconstruction in these regions and higher MAE. These observations suggest that while AnyECG excels in preserving overall rhythmic and morphological integrity, there is room for improvement in accurately reconstructing fine-grained details.

\begin{table}{}
  \centering
  \caption{Results Comparison with State-of-the-Art Models in Corrupted Lead Generation.}
  \label{tab:Generation}
  \adjustbox{max width=0.60\textwidth}{
    \begin{tabular}{lccc}
    \toprule
    \textbf{Methods} & \textbf{PSNR} $\uparrow$ & \textbf{SSIM} $\uparrow$ & \textbf{MAE} $\downarrow$  \\
    \midrule
    CGAN~\cite{mirza2014conditional} & 30.1762 &  0.8591 & \textbf{0.0142}  \\
    WGAN~\cite{adler2018banach} &  27.5074 & 0.7907 & 0.0199  \\
    % Nef-Net~\cite{chen2021electrocardio} & 15.0892 & 0.8672 & 0.8672 \\
    % ECG-FM~\cite{mckeen2024ecg} & - & - & -  \\
    \midrule
    \model-B & 32.5456 & 0.8634 & 0.0312  \\
    \model-L & \textbf{32.7372} & \textbf{0.8738} &  0.0296 \\
    \model-XL & 32.4276 & 0.8529 & 0.0376  \\
    \bottomrule
    \end{tabular}
  }
\vskip -10pt
\end{table}

\paragraph{Ultra-Long ECG Analysis}
Recognizing ultra-long ECG signals is challenging due to their extended duration, rhythm variability, and noise, which require models to be robust and generalizable. Traditional time series models often struggle with high computational complexity, memory constraints, and difficulty in capturing long-term dependencies. Therefore, We proposed a hierarchical modeling approach that adapts to ultra-long ECG data by employing a sliding window method. As shown in Table \ref{tab:long}, \model, particularly the \model-XL, achieves the highest scores across all evaluation metrics. This demonstrates its superior ability to capture complex patterns and maintain high accuracy when analyzing ultra-long ECG signals. Compared to state-of-the-art models like Inception1D~\cite{strodthoff2020deep} and RNN1D~\cite{salloum2017ecg}, \model-XL shows a clear advantage, especially in AUROC and AUC-PR. Even the smaller variants, \model-B and \model-L, outperform most baseline models, highlighting \model's adaptability and scalability. The absence of results for the other pretrained ECG foundation model ECG-FM~\cite{mckeen2024ecg} is due to its inability to handle ultra-long sequence data, making it unsuitable for this task. In contrast, \model's consistent performance across all scales confirms its effectiveness in capturing key features of ultra-long ECG signals.
% \subsection{SCALING DATA SIZE}

% While six datasets of ECG data were collected for pre-training, further scaling of data size remains a critical exploration area. Preliminary findings suggest that performance scales with data volume, particularly for larger models like \model-Huge, which continue to benefit from increased training data sizes. Future work will investigate the data size requirements to further enhance model performance, particularly for larger-scale ECG tasks.

% \begin{figure*}[t]
%     \centering
%     \includegraphics[width=0.99\textwidth]{ECGFM@ICLR25/fig/Generation.pdf}
%     \caption{\textbf{Visualization of Corrupted Lead Generation}
%     \label{fig:generation}
% \end{figure*}

% \begin{figure}
%     \centering
 
%     \begin{subfigure}{\textwidth}
%         \centering
%         \includegraphics[width=\textwidth]{ECGFM@ICLR25/fig/WGAN.pdf}
%         \label{fig:sub1}
%     \end{subfigure}
%     % \hfill
%     \begin{subfigure}{\textwidth}
%         \centering
%         \includegraphics[width=\textwidth]{ECGFM@ICLR25/fig/WGAN.pdf}
%         \label{fig:sub2}
%     \end{subfigure}
%     % \hfill
%     \begin{subfigure}{\textwidth}
%         \centering
%         \includegraphics[width=\textwidth]{ECGFM@ICLR25/fig/WGAN.pdf}

%         \label{fig:sub3}
%     \end{subfigure}
    
%     \caption{picture}
%     \label{fig:three_subfigures}
% \end{figure}

\subsection{Pre-training Phase Ablation Study}
\label{sec:ablation_pretrain}

\begin{figure}[htbp]
    \centering
    \begin{minipage}{0.49\textwidth}
        \centering
        \includegraphics[width=\textwidth]{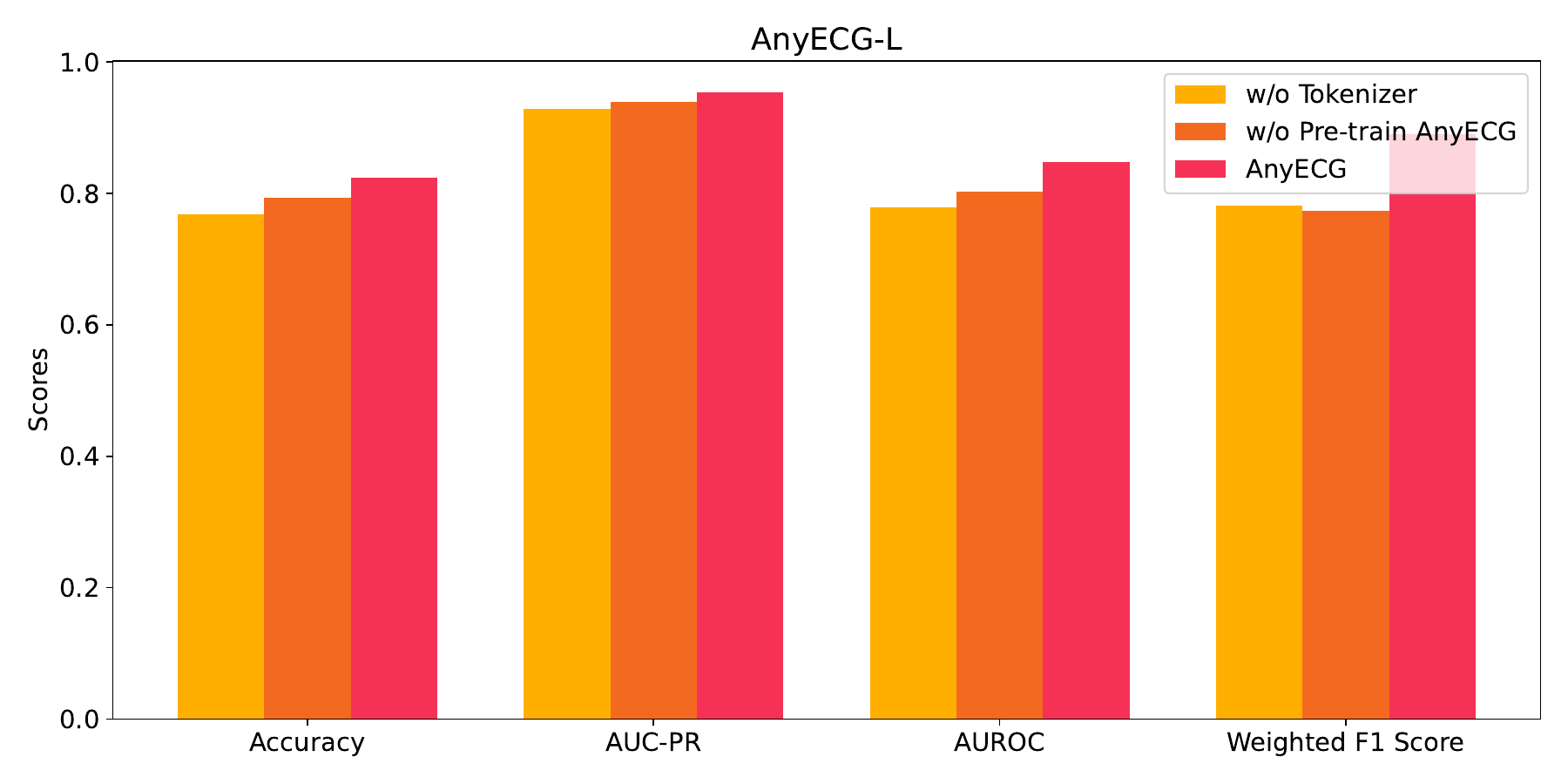}
    \end{minipage}
    \hfill
    \begin{minipage}{0.49\textwidth}
        \centering
        \includegraphics[width=\textwidth]{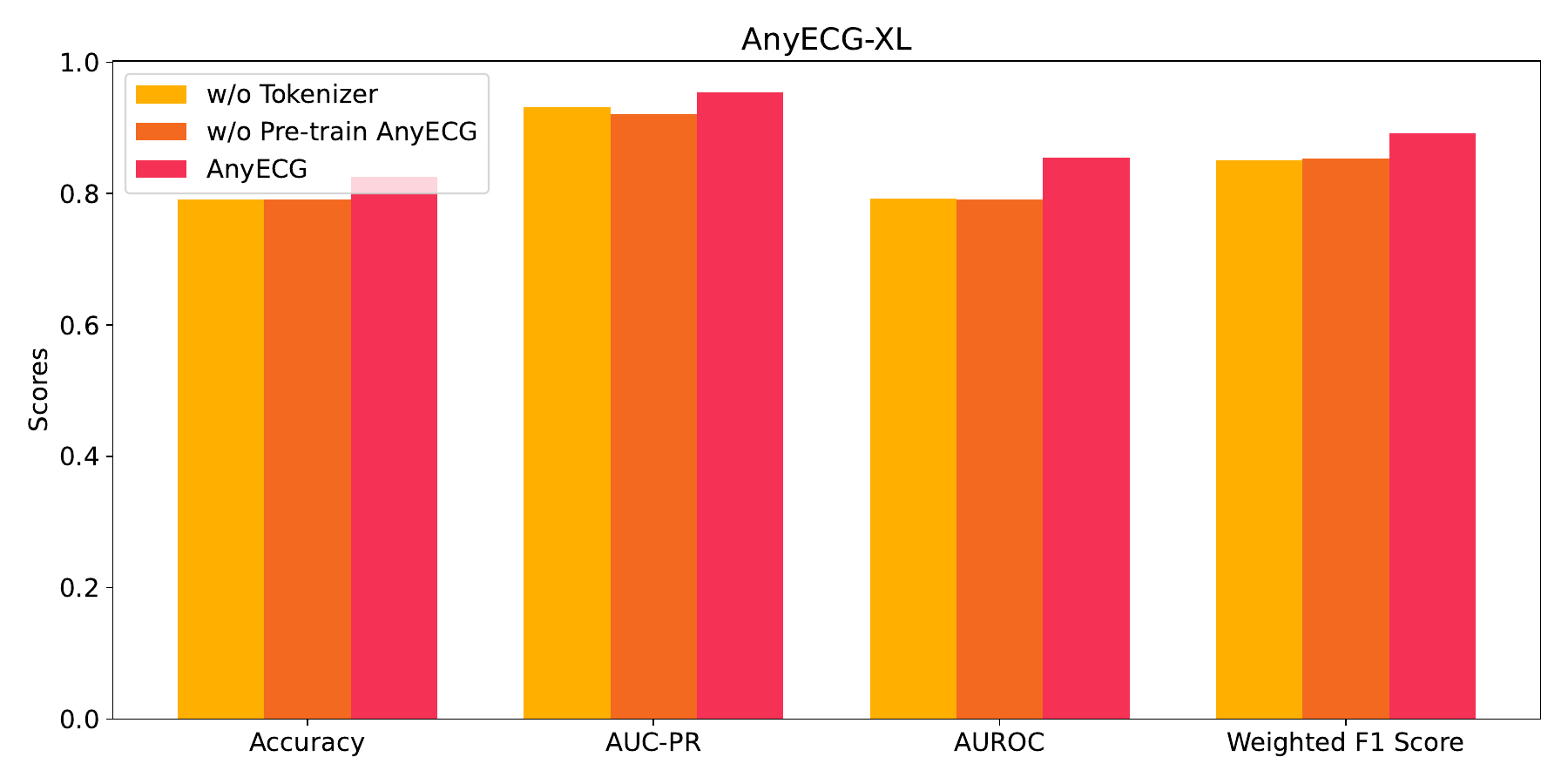}
    \end{minipage}
    \caption{Ablation study of Pre-training Phase in Anomaly Detection with AnyECG-L and AnyECG-XL}
    \label{fig:ablation}
\end{figure}

To evaluate the contribution of each component in our pre-training strategy, we conducted an ablation study focusing on the pre-training phases. Specifically, we analyzed the effects of pre-training the {\tokenizer} and the full {\model} foundation model on anomaly detection performance. Table~\ref{tab:Pre-training——Ablation} and Figure~\ref{fig:ablation} presents the results of this study. The first configuration is {\model-B} without {\tokenizer} pre-training. The second configuration includes a pre-trained {\tokenizer} but skips pre-training the {\model} foundation model. The final configuration involves full pre-training of both the {\tokenizer} and the {\model}. The results show that pre-training the {\tokenizer} leads to noticeable improvements over the baseline.  This indicates that a pre-trained {\tokenizer} enhances the model's ability to capture meaningful representations of the ECG signals. When the full {\model} foundation model is also pre-trained, we observe a significant performance boost across all metrics. These gains underscore the importance of comprehensive pre-training in enhancing the model's anomaly detection capabilities. The fact that full pre-training yields the best results confirms that both components—the {\tokenizer} and the {\model}—contribute positively to the overall performance. Pre-training the {\model} foundation model allows it to learn generalizable features that are beneficial for downstream tasks, while the pre-trained {\tokenizer} ensures effective encoding of the input signals.

\begin{table*}[h]
\centering
\caption{Ablation study of Pre-training Phase in Anomaly Detection}
\label{tab:Pre-training——Ablation}
\adjustbox{max width=\textwidth}{
\begin{tabular}{lcccc}
\toprule
\textbf{Methods}  & \textbf{Accuracy} $\uparrow$ & \textbf{AUC-PR} $\uparrow$ & \textbf{AUROC} $\uparrow$ & \textbf{Weighted F1 Score} $\uparrow$ \\
\midrule

{\model-B}  & \multirow{2}{*}{0.7623} & \multirow{2}{*}{0.9243} & \multirow{2}{*}{0.7729} & \multirow{2}{*}{0.7512} \\
w/o \tokenizer & & & & \\  
\midrule
{\model-B}  & \multirow{2}{*}{0.7810} & \multirow{2}{*}{0.9358} & \multirow{2}{*}{0.8232} & \multirow{2}{*}{0.7826} \\
w/o Pre-train \model &  & & & \\  
\midrule
\multirow{2}{*}{\model-B} & \multirow{2}{*}{\textbf{0.8188}} & \multirow{2}{*}{\textbf{0.9517}} & \multirow{2}{*}{\textbf{0.8502}} & \multirow{2}{*}{\textbf{0.8863}} \\
 & & & & \\  

\bottomrule
\end{tabular}
}
\end{table*}

\begin{table*}[!h]
\centering
\caption{Ablation Study on Loss Function Components in Anomaly Detection}
\label{tab:loss_ablation}
\adjustbox{max width=\textwidth}{
\begin{tabular}{lcccc}
\toprule
\textbf{Loss Configuration} & \textbf{Accuracy} $\uparrow$ & \textbf{AUC-PR} $\uparrow$ & \textbf{AUROC} $\uparrow$ & \textbf{Weighted F1 Score} $\uparrow$ \\
\midrule
Full Loss & \textbf{0.8188} & \textbf{0.9517} & \textbf{0.8502} & \textbf{0.8863} \\
w/o Morphology Loss & 0.8059 & 0.9373 & 0.8381 & 0.8754 \\
w/o Frequency Loss & 0.8125 & 0.9412 & 0.8475 & 0.8621 \\
w/o Demography Loss & 0.8134 & 0.9487 & 0.8445 & 0.8801 \\
w/o Codebook Loss & 0.7522 & 0.8950 & 0.7900 & 0.8150\\
w/o Commitment Loss & 0.7855 & 0.9275 & 0.8225 & 0.8575 \\
\bottomrule
\end{tabular}
}
\end{table*}

\subsection{Loss Function Ablation Study}
\label{sec:ablation_loss}

To assess the effectiveness of our loss function design, we conducted an ablation study by systematically removing each component of the loss function. This allowed us to evaluate how each term contributes to the model's ability to capture meaningful features from ECG signals. Table~\ref{tab:loss_ablation} presents the results of this study. The "Full Loss" configuration, which includes all components of our proposed loss function, serves as the baseline for optimal performance. When we individually removed each loss component, we observed a decrease in performance across all evaluation metrics. Omitting the Morphology Loss resulted in a noticeable decline, indicating its significant role in helping the model capture the morphological characteristics of ECG signals, which are crucial for accurate anomaly detection. Excluding the Frequency Loss also led to reduced performance, suggesting that capturing frequency domain information is important for understanding underlying patterns in ECG signals. Removing the Demography Loss caused a performance drop as well, though to a lesser extent compared to the Morphology and Frequency losses. This highlights that incorporating demographic information refines the model's predictions by accounting for variations in ECG patterns across different demographic groups. The most significant decrease in performance was observed when the Codebook Loss was removed. This component is essential for encouraging diversity and utilization of codebook entries in the vector quantization process, playing a critical role in the model's ability to represent ECG signals effectively. Lastly, removing the Commitment Loss also led to a decline in performance, though the impact was less severe than omitting the Codebook Loss. The Commitment Loss ensures consistency in the representation of similar inputs by encouraging the encoder to commit to specific codebook entries. The combination of Morphology, Frequency, Demography, Codebook, and Commitment losses enables the model to capture comprehensive features of ECG signals, leading to improved anomaly detection capabilities. These results validate the design of our loss function and underscore the importance of each term in capturing different aspects of the ECG data.

% \begin{table}[!h]
% \centering
% \caption{Ablation Study on Loss Function Components in Anomaly Detection}
% \label{tab:loss_ablation}
% \adjustbox{max width=\textwidth}{
% \begin{tabular}{lcccc}
% \toprule
% \textbf{Loss Configuration} & \textbf{Accuracy} $\uparrow$ & \textbf{AUC-PR} $\uparrow$ & \textbf{AUROC} $\uparrow$ & \textbf{Weighted F1 Score} $\uparrow$ \\
% \midrule
% Full Loss & \textbf{0.8188} & \textbf{0.9517} & \textbf{0.8502} & \textbf{0.8863} \\
% w/o Morphology Loss & 0.8059 & 0.9373 & 0.8381 & 0.8754 \\
% w/o Frequency Loss & 0.8125 & 0.9412 & 0.8475 & 0.8621 \\
% w/o Demography Loss & 0.8134 & 0.9487 & 0.8445 & 0.8801 \\
% w/o Codebook Loss & 0.7522 & 0.8950 & 0.7900 & 0.8150\\
% w/o Commitment Loss & 0.7855 & 0.9275 & 0.8225 & 0.8575 \\
% \bottomrule
% \end{tabular}
% }
% \end{table}

% \subsection{Visualization of ECG Leads Generation}

\vspace{0.5cm}

%% file: sec/7.discussion.tex
\section{Discussion}

\paragraph{Social Impacts}
ECG is one of the most commonly used diagnostic tools in healthcare, with over 100 million ECG reports obtained annually in the United States alone~\cite{tison2019automated}. Despite its widespread use, unlike other biomedical signals such as electroencephalograms (EEG)~\cite{yang2024biot,jiang2024large}, there is a scarcity of foundation models specifically designed for ECG data. This limitation hampers the potential for advanced analysis and interpretation of ECG signals on a large scale. In this work, we propose {\model}, the largest ECG foundation model family to date. Compared to prior works~\cite{mckeen2024ecg,song2024foundation,fu2024cardiogpt}, {\model} adapts to diverse downstream tasks and achieves significantly better performance. By providing a robust and generalizable model for ECG data, {\model} has the potential to greatly enhance diagnostic accuracy, facilitate early detection of cardiovascular diseases, and improve patient outcomes on a broad scale.

\paragraph{Limitations}
Although we pre-trained {\model} using a large amount of data across seven datasets, there remains a significant gap between {\model} and current foundation models like LLMs in the general domain. This gap is primarily due to the difficulty in obtaining extensive healthcare data. Additionally, the model size of {\model}-XL (1.7B parameters) is considerably smaller than that of foundation models in natural language processing and computer vision fields. Despite these limitations, it is important to highlight that training a large-scale ECG foundation model with a two-stage self-supervised learning approach and more data does yield appreciable performance gains compared to existing methods developed for specific downstream tasks, even if it may be computationally costly. Exploring the trade-off between employing larger {\model} models and enhancing downstream task performance will be a focus of our future work.

%% file: sec/6.conclusion.tex
\section{Conclusion}

In this paper, we proposed {\model}, a foundation model family that learns universal embeddings through a two-stage self-supervised pre-training on seven diverse ECG datasets. {\model} effectively handles the heterogeneity of ECG data through the design of a novel Rhythm Quantizer, which includes a rhythm codebook and a multi-view synergistic decoder to learn representations from different proxy tasks. Additionally, masked modeling in the second-stage pre-training plays a crucial role in enabling effective representation learning of both temporal and lead features of ECG signals. We validated various sizes of {\model} models on multiple downstream tasks, including anomaly detection, arrhythmia detection, ECG lead generation, and ultra-long ECG signal recognition. Our experiments demonstrate that {\model} outperforms all state-of-the-art methods in their respective fields, highlighting its effectiveness and versatility in ECG signal analysis.

%% file: sec/appendix.tex
\subsection{Notations}
% \label{sec:appx_notation}

% \xunote{Update all notations here:}
% \twocolumn

\vspace{0.5cm}
\begin{supertabular}{p{3cm} p{5cm}}
    \hline
    \multicolumn{2}{l}{\textbf{Data and Indices}} \\
    \hline
$X \in \mathbb{R}^{L \times T}$ & Multi-channel ECG signals \\
$L$ & Number of ECG leads \\
$T$ & Total time steps \\
$s$ & Patch size (time steps) \\
$P$ & Number of patches per lead \\
$x_{j,k} \in \mathbb{R}^{s}$ & Patch from lead $j$, index $k$ \\
$N = L \times P$ & Total number of patches \\
$j = 1, 2, \dotsc, L$ & Lead index \\
$k = 1, 2, \dotsc, P$ & Patch index \\
\midrule
\multicolumn{2}{l}{\textbf{Embeddings and Positional Encodings}} \\
\midrule
$d$ & Embedding dimension \\
$h'_{j,k} \in \mathbb{R}^d$ & Patch embedding \\
$\tau_k \in \mathbb{R}^d$ & Temporal positional encoding \\
$\sigma_j \in \mathbb{R}^d$ & lead positional encoding \\
${h}_{j,k} = h'_{j,k} + \tau_k + \sigma_j$ & Augmented embedding \\
\midrule
\multicolumn{2}{l}{\textbf{Neural Tokenizer and Codebook}} \\
\midrule
$V \in \mathbb{R}^{K \times d}$ & Codebook of codewords \\
$i^\ast$ & Codeword index \\
$z_{j,k} = i^\ast$ & Discrete token \\
$K$ & Number of codewords \\
\midrule
\multicolumn{2}{l}{\textbf{Attention Mechanism and Transformer Components}} \\
\midrule
$H \in \mathbb{R}^{N \times d}$ & Embedding matrix \\
$Q$, $K$, $V$ & Query, key, value matrices \\
$W_Q$, $W_K$, $W_V \in \mathbb{R}^{d \times d_{\text{model}}}$ & Projection matrices \\
$\text{LayerNorm}(\cdot)$ & Layer normalization \\
$M \in \mathbb{R}^{N \times N}$ & Attention mask \\
$\mathcal{A}(i)$ & Attention set for patch $i$ \\
$d_{\text{head}}$ & Head dimension \\
$d_{\text{model}}$ & Model dimension \\
\midrule
\multicolumn{2}{l}{\textbf{Decoders and Reconstruction}} \\
\midrule
$o_{j,k}^m$ & Morphology decoder output \\
$c_{A}^{(l)}[n]$ & Approximation coefficients \\
$c_{D}^{(l)}[n]$ & Detail coefficients \\
$c_{A}^{(l)\, \text{norm}}$, $c_{D}^{(l)\, \text{norm}}$ & Normalized coefficients \\
$\hat{c}_{A}^{(l)}$, $\hat{c}_{D}^{(l)}$ & Predicted coefficients \\
$g[\cdot]$, $h[\cdot]$ & Filter coefficients \\
$a \in \mathbb{R}^{d_a}$ & Demography vector \\
$o^a$ & Predicted demography \\
$\hat{x}_{j,k}$ & Reconstructed patch \\
$f$ & Decoder function \\
\midrule
\multicolumn{2}{l}{\textbf{Loss Functions}} \\
\midrule
$\mathcal{L}_{\text{morphology}}$ & Morphology loss \\
$\mathcal{L}_{\text{freq}}$ & Frequency loss \\
$\mathcal{L}_{\text{demography}}$ & Demography loss \\
$\mathcal{L}_{\text{codebook}}$ & Codebook loss \\
$\mathcal{L}_{\text{commitment}}$ & Commitment loss \\
$\mathcal{L}_T$ & Total tokenizer loss \\
$\mathcal{L}_{\text{mask}}$ & Masked modeling loss \\
% $\mathcal{L}_{\text{mar}}$ & Morphological autoregression loss \\
% $\mathcal{L} = \lambda_{\text{mask}} \mathcal{L}_{\text{mask}} + \lambda_{\text{mar}} \mathcal{L}_{\text{mar}}$ & Overall loss \\
% $\beta$, $\lambda_{\text{mask}}$, $\lambda_{\text{mar}}$ & Weighting coefficients \\
\midrule
\multicolumn{2}{l}{\textbf{Masking and Autoregression}} \\
\midrule
$m_{j,k} \in \{0, 1\}$ & Mask indicator \\
$h_M \in \mathbb{R}^d$ & Mask token \\
$\tilde{h}_{j,k}$ & Masked embedding \\
$\bar{h}_{j,k} = \tilde{h}_{j,k} + \tau_k + \sigma_j$ & Augmented masked embedding \\
$\tilde{h}'_{j,k}$ & Contextualized representation \\
$p(v_i \mid \tilde{H})$ & Probability distribution \\
\midrule
\multicolumn{2}{l}{\textbf{Other Parameters and Hyperparameters}} \\
\midrule
$L_w$ & Decomposition levels \\
$d_a$ & Demography vector dimension \\
$n$ & Coefficient index \\
$r$ & Masking ratio \\
    \hline
\end{supertabular}